\documentclass[lettersize,journal]{IEEEtran}
\usepackage{amsmath,amsfonts}
\usepackage{algorithmic}
\usepackage{algorithm}
\usepackage{array}
\usepackage[caption=false,font=normalsize,labelfont=sf,textfont=sf]{subfig}
\usepackage{textcomp}
\usepackage[dvipsnames]{xcolor}
\usepackage{stfloats}
\usepackage{authblk}
\usepackage{url}
\usepackage{hyperref}
\usepackage{verbatim}
\usepackage{graphicx}
\usepackage{cite}

\IEEEoverridecommandlockouts
\IEEEpubid{
\begin{minipage}{\textwidth}\ \\[40pt]
        \centering\normalsize{This work has been submitted to the IEEE for possible publication. Copyright may be transferred without notice, after which this version may no longer be accessible.}
\end{minipage}
}
\begin{document}

\title{Heterogeneity- and homophily-induced vulnerability of a P2P network formation model: the IOTA auto-peering protocol}

\author[1]{Yu Gao}
\author[2,3,5]{Carlo Campajola}
\author[1,2,*]{Nicol\`{o} Vallarano}
\author[4]{Andreia Sofia Teixeira}
\author[1,2]{Claudio J. Tessone}

\affil[1]{Blockchain \& Distributed Ledger Technologies Group, Department of Informatics, University of Z\"{u}rich, Switzerland}
\affil[2]{UZH Blockchain Center, University of Z\"{u}rich, Switzerland}
\affil[3]{Institute of Finance and Technology, University College London, United Kingdom}
\affil[4]{LASIGE, Departamento de Informática, Faculdade de Ciências, Universidade de Lisboa, Portugal}
\affil[5]{DLT Science Foundation, London, United Kingdom}
\affil[*]{\small Corresponding author: nicolo.vallarano@uzh.ch}


\maketitle
\IEEEpubidadjcol
\begin{abstract}
IOTA is a distributed ledger technology that relies on a peer-to-peer (P2P) network for communications. Recently an auto-peering algorithm was proposed to build connections among IOTA peers according to their ``Mana" endowment, which is an IOTA internal reputation system. This paper's goal is to detect potential vulnerabilities and evaluate the resilience of the P2P network generated using IOTA auto-peering algorithm against eclipse attacks.
In order to do so, we interpret IOTA's auto-peering algorithm as a random network formation model and employ different network metrics to identify cost-efficient partitions of the network.
As a result, we present a potential strategy that an attacker can use to eclipse a significant part of the network, providing estimates of costs and potential damage caused by the attack.
On the side, we provide an analysis of the properties of IOTA auto-peering network ensemble, as an interesting class of homophile random networks in between 1D lattices and regular Poisson graphs.
\end{abstract}





\section{Introduction}

\IEEEPARstart{O}{ver} the last few decades peer-to-peer networks (P2P) have been widely employed in a variety of settings as a base layer for distributed systems, including the worldwide web \cite{lawrence2000accessibility}, file sharing \cite{oram2001peer}, instant messaging in social networks \cite{i2001topology,heidemann2012online,masinde2020peer}, and distributed computing \cite{tanenbaum2007distributed}. In contrast to the client-server system, P2P networks stand out as more egalitarian and decentralised, where peers share data resources with each other without resorting to a central authoritative node \cite{androutsellis2004survey}. Particularly, the emergence of Bitcoin \cite{nakamoto2008bitcoin} has brought revolutionary technology in socio-technical and socio-economic systems, and the decentralised P2P network layer is the backbone of its Distributed Ledger Technology (DLT).

Blockchains and Distributed Ledger Technologies have gained tremendous traction in the last decade, and advances pervade socio-technical and socio-economic aspects surrounding them. A DLT is a consensus-driven system where digital data is replicated, shared, and synchronised across multiple sites, countries, or institutions in the absence of a central authority, ensuring geographic decentralisation. 
A blockchain is a specific type of DLT, with Bitcoin \cite{nakamoto2008bitcoin}, the first type of a DLT to reach worldwide diffusion, using a blockchain for transaction storing.

Generally, the P2P network in DLTs is the infrastructure that enables the distribution of data among peers, and the establishment of consensus is significantly influenced by the topology of connections within the P2P network \cite{el2018review,kraner2023agent}. The nodes in a generic P2P network are computers, and the information they contain and exchange can be files, contracts, and transaction records, to name a few \cite{wood2014ethereum}. To give an example, if we think of DLTs storing specifically economic transactions like Bitcoin, the ``miners" i.e. the peers driving consensus play the role of clerks in a bank: they clear the systems transactions in order for the everyday Bitcoin user to be able to execute their transactions. Their incentive in doing so is the coinbase reward associated with each block attached to the blockchain, which also serves as a Bitcoin minting mechanism.

Due to the growing adoption of cryptocurrencies, numerous researchers have extensively explored their P2P networks. One of the main challenges in studying real-world P2P networks follows from their decentralised nature: they rarely can be fully monitored by a single agent, as each node only knows about its peers and has no reliable information beyond them, nor are they incentivised to share this information as it may lead to malicious attacks. To address this problem, several scholars have proposed methodologies for mapping these networks and conducting subsequent analyses. To cite a few, Miller introduced ``AddressProbe" to uncover the topology of the Bitcoin P2P network \cite{miller2015discovering}; Deshpande et al. devised a framework to map the Bitcoin P2P network, presenting an insightful snapshot of its topology \cite{deshpande2018btcmap}; Kim et al. developed NodeFinder, a tool to measure the Ethereum P2P network's characteristics and dissect the intricate DEVp2p ecosystem \cite{kim2018measuring}. From a different viewpoint, Imtiaz et al. revealed the intermittent network connectivity displayed by Bitcoin nodes \cite{imtiaz2019churn}.

Notably, there exists research concerning attacks on P2P networks: Heilman et al. \cite{heilman2015eclipse} proposed employing extremely low-rate TCP connections for network attacks, while Nayak et al. \cite{nayak2016stubborn} devised a combination of mining attacks and network-level attacks to undermine the consensus in DLT protocols. Additionally, other papers outline a variety of technical approaches to network attacks \cite{apostolaki2017hijacking,tran2020stealthier}.

In P2P networks, nodes actively communicate with nearby nodes, facilitating the exchange of information. This dynamics is crucial for maintaining consensus among nodes and deterring potential efforts by individuals or attackers to gain complete control of the network. Despite the original intention of most DLT systems' P2P networks to build random, uncontrollable networks, vulnerabilities can arise when certain network information is exploited \cite{muller2021salt,moubarak2018blockchain}.
To prevent such attacks, it is essential to have a comprehensive understanding of the inherent topology of the empirical P2P network within a static network model, achieved by substituting technical cryptographic elements with graph-based parameters \cite{dobson2007complex, arianos2009power,epstein2012generative,hines2010topological}.
This approach can uncover potential weaknesses in the design of P2P network protocols.

Network Science offers a comprehensive tool-set to effectively model these types of complex systems \cite{newman2018networks,watts1998collective}. Socio-technical systems have been largely studied in the context of complex networks \cite{vespignani2012modelling,watts2005multiscale}: translating such systems into network representations allows us to exploit their topological properties in order to identify vulnerabilities and nonlinear properties. P2P protocols have a natural network representation, where peers can be represented as nodes, while edges embody peer connections. An obstacle to a straightforward application of network analysis to these systems is that by design P2P protocols make the connection construction unknown globally, as peers only know their local connectivity. On the contrary, peers generally have a huge security incentive not to reveal their connections to possible malicious adversaries \cite{heilman2015eclipse,singh2006eclipse}.
A possible approach to this challenge is to define theoretical models that describe P2P networks, starting from the P2P protocol rules, in order to gain further insight.

The prevailing research naturally focuses on Bitcoin and Ethereum, both operating in a permissionless manner. Generally, these platforms incentivise validators — i.e. the peers tasked with verifying transactions and upholding network integrity — by issuing rewards for their honest contribution and penalties if their actions contradict rules or network interests. However, for platforms lacking such reward and penalty systems for peers like IOTA\footnote{\href{https://www.iota.org/} {https://www.iota.org/}}, the design of the peer connection algorithms can have a profound influence on the consensus integrity.

IOTA has been introduced to address one of the main limitations of mainstream blockchains, namely their scalability. Blockchain addresses the challenge of decentralising digital transaction systems while enhancing data transparency and security. However, it may encounter scalability issues when managing a large volume of transactions \cite{khan2021systematic}. Each node is required to verify and store a complete copy of the blockchain, leading to limited network capacity and slower transaction processing; in addition to this, blockchains have an intrinsic transaction volume limit, depending on how many transactions can fit inside a block and the time that occurs between two blocks. To tackle this, IOTA introduced the Tangle \cite{popov2018tangle}, a new ledger data structure based on a directed acyclic graph (DAG) instead of the classic linear chain. By enabling concurrent transaction processing, the Tangle can theoretically handle increased transaction throughput as the network expands.
Another peculiarity of IOTA is  ``\textit{Mana}", a scarce quantity associated with a peer that would tokenize its reputation, which has been proposed to be introduced in an upcoming upgrade \cite{popov2020coordicide}. Mana is intended as a Sybil protection mechanism \cite{douceur2002sybil, cheng2005sybilproof}, i.e. a mechanism that sets a cost (in this case, acquiring reputation) for nodes to participate in the consensus formation. This prevents Sybil attacks, where malicious agents are able to manipulate consensus by the creation of a large number of nodes. 

It was proposed to use Mana within a pseudo-random P2P network formation model, the IOTA auto-peering module \cite{muller2021salt}, which favours homophilic connections, i.e. among nodes in a close Mana range. 
Homophily is known in sociology as the tendency of individuals to bind by similarities. In network theory, we observe homophily when nodes tend to connect preferentially with nodes with similar characteristics \cite{newman2002assortative,kossinets2009origins,shalizi2011homophily,hasheminezhad2023robustness}.
Many social networks demonstrate connection mechanisms in which each node possesses a distinct trait, such as wealth, age, or trustworthiness. Connections in these networks arise from similarities, as observed in friendship networks \cite{moody2001race},  collaboration networks \cite{newman2001structure} or actor networks \cite{watts1999networks}.

The goal of our work is to identify potential weaknesses in the P2P network formation model that may result in vulnerability to attacks. In particular, we focus on eclipse and partitioning attacks, where malicious actors aim at controlling the flow of information between portions of the network to an extent where they gain an unfair amount of power in determining the consensus. To investigate the security of the IOTA P2P network, we analyse the networks generated by the IOTA auto-peering formation model \footnote{\url{https://github.com/iotaledger/autopeering-sim}\label{fn:source}} 
and develop two attack strategies: the ``Betweenness" strategy and the ``Greedy" strategy. From an attacker's perspective, both strategies can be translated into network partitioning algorithms: what the attacker is interested in is to find the ``control set", i.e. the subset of nodes they need to take control of in order to censor the flow of information between the two network partitions.
We initially employ the ``Betweenness" strategy, which utilises the link betweenness centrality metric \cite{barthelemy2004betweenness,brandes2001faster, brandes2008variants} to split the P2P network by removing the top betweenness links and by identifying the control set as the nodes that are endpoints of the removed links. The second strategy we employ, which we call ``Greedy", involves sorting all nodes in the network by Mana endowment and defining a possible partition by splitting nodes according to their ranking. To quantify the impact of the attack, we introduce the concepts of ``Damage" and ``Cost", whose ratio we employ as a measure of attack efficiency.


Although both strategies rely on complete topological information about the network, it is important to note that in real-world scenarios the specific P2P network topology is not publicly accessible. Therefore, to address this limitation, we also propose a ``Blind" strategy. This strategy is inspired by the results of the ``Betweenness" and ``Greedy" strategies, but it is executed assuming the attacker has no knowledge of the actual topology of the P2P network except for the knowledge of the Mana distribution and the IOTA auto-peering network formation model parameters, which are public information available to all IOTA nodes.
The blind strategy identifies a ``target" node, i.e. the centre of the control set, which is then composed of a specific number of nodes before and after the target node in terms of Mana ranking. Our experiments show that blind strategies can be applied to obtain advantageous control sets in terms of attack efficiency, leading to the P2P network partition into two separate components that causes more damage than the attack cost. We also test the performances of blind strategy attacks on networks generated using the Watts-Strogatz (WS) model \cite{watts1999networks, watts1998collective}, in order to have a reference baseline. Indeed, as will be clear from the results, the WS networks ranging from the 1D lattice to a fully randomised regular network act as the ideal reference for the IOTA auto-peering networks. These are in fact themselves random regular networks, with a preferential dimension for connections given by the Mana endowment rather than by the position in the 1D lattice.

The rest of the paper is organised as follows: Section II describes our approach to the problem, setting the theoretical framework for the simulation study. Section III presents the results of the simulations. Section IV concludes the paper.

\section{Approach}

\subsection{Mana}

The \textit{auto-peering module}, as described by IOTA\textsuperscript{\ref{fn:source}},
generates a P2P network composed of $N$ nodes. Each node has a given amount of Mana and, for convenience, we will index them from the largest endowment of Mana (identified as $1$) to the lowest (identified as $N$). These Mana values are publicly known to all nodes and regularly broadcast over the P2P network itself. The distribution of Mana is the result of a dynamical process, where nodes can accrue their endowment through fair participation in the consensus mechanism, and their chance to do so is itself a function of their Mana endowment; in our setting we assume Mana to be an equilibrium value, not depending on time. As is commonly found in similar systems like Proof-of-Work mining \cite{campajola2022evolution, makarov2021blockchain}, Proof-of-Stake validation \cite{chohan2022cryptocurrencies, sai2021characterizing} or any system where the growth of a quantity is proportional to its current level, the so-called ``Matthew effect" enters the picture and a Zipf's law often emerges after sufficient time. For this reason, we assume the Mana distribution follows a Zipf's Law \cite{zipf2016human} with exponent $s$ so that the Mana value of node $i$ is

\begin{equation}\label{q:1}
    m(i) = K i^{-s},
\end{equation}
 
where $m(i)$ is the Mana endowment of the $i$-th ranked node and $K$ is an arbitrary positive constant used for numeric stability, in our case $K=10^{10}$.

\subsection{Peer discovery}
The protocol prescribes that each node $i$ will add node $j > i$ to its list of eligible neighbours, $\mathcal{N}_i$, if $m(i)<\rho m(j)$, where $\rho$ is a given constant. If this condition is not met for at least $R$ other nodes, $i$ will add the $R$ nodes ranked after itself to the list. Any time a node $j$ is added to the set $\mathcal{N}_i$, $i$ will be added to $\mathcal{N}_j$ to reciprocate.

Therefore, the potential neighbours of any node $i$ are formally identified as

\begin{equation}\label{q:2}
\small
\mathcal{N}_i = \left\lbrace j = 1, \dots, N \, : \, \frac{1}{\rho} m(i) < m(j) < \rho m(i) \, \vee \, \left\vert j-i \right\vert < R \right\rbrace .
\end{equation}
From this set, $i$ randomly picks $k$ elements to link to and receives $k$ links from other nodes in the set. The result is a random regular network with coordination number $k$ (or in other words where each node has degree $2k$), and links appear only between nodes with similar Mana endowment. From a modelling perspective, $N$, $k$, $R$, $\rho$ and the Zipf's exponent $s$ are the parameters needed to generate a network.


\subsection{Network Formation}\label{sec:net-model}

We base our simulation on the publicly available Go code by the IOTA research team\footnote{\url{https://github.com/iotaledger/autopeering-sim}\label{fn:source}}. A node will compile its potential neighbours' list and cease searching for neighbours when it has established $k$ connections, while it accepts connection requests from other nodes until it has received another $k$ connections. Communication is then bidirectional along these links, so the resulting network is undirected. As a result, each node ultimately has $2k$ neighbours.


Combining the aforementioned rules, we can deduce that an increase in the Zipf's exponent $s$ leads to the most affluent nodes having relatively short potential neighbour lists. This happens because of the increased inequality in the distribution of Mana among nodes, meaning a larger Mana gap between nodes.
The parameters $R$ and $\rho$, although not as influential, still impact the IOTA P2P network and contribute to its structure by tuning the tolerance that the nodes have in terms of the Mana endowment of their potential neighbours. As can be seen by the definition of Eqs. \ref{q:1} and \ref{q:2}, when $s$ is small the distribution is relatively homogeneous and peers easily have many potential neighbours through the range condition set by $\rho$, forming a network that is closely resembling a regular (i.e. where all nodes have the same degree) version of the Erd\H{o}s-R\'{e}nyi (ER) random network. However, as $s$ increases, nodes that fail to find enough peers within the Mana ratio condition can acquire sufficient potential neighbours through the alternative condition based on $R$.   In this situation, peers that have fewer potential neighbours can solely establish connections with peers that are in their immediate vicinity in terms of Mana ranking. Consequently, the network will tend towards resembling a k-regular 1D lattice network.

\begin{figure*}[!t]
\centering
    \includegraphics[ width=0.8\textwidth]{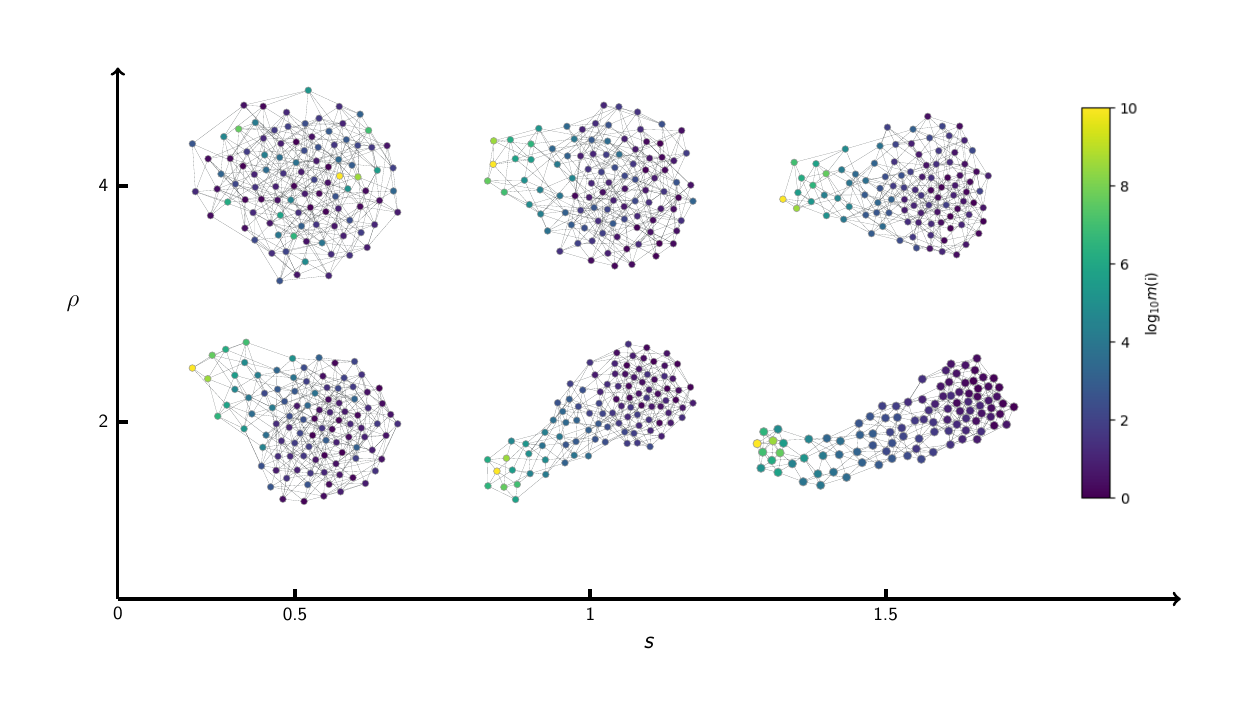}
    \caption{Intuitive visualisation of the different topologies realised by the IOTA auto-peering formation model as a function of $\rho$ and $s$. The graphs are generated with fixed $N=100, R=10$, $k=4$. Colour represents the Mana endowment of nodes, in a logarithmic scale.}
    \label{fig:log-Mana}
\end{figure*}

     
To have a more intuitive understanding of the IOTA auto-peering model, we provide a visualisation of networks generated under various $s$ and $\rho$ values in Figure \ref{fig:log-Mana}. As it can be seen, when the Mana (shown in colour in log scale) is relatively homogeneous and when the tolerance parameter $\rho$ is high enough the network does not show any particular asymmetry, but in a low $\rho$ scenario and whenever $s$ is increased there is a clear orientation of the network, with the nodes with the largest Mana endowment being shifted towards one end of the network and the shape becomes more linear, following the nodes Mana hierarchy.

\subsection{Attack Damage and Cost}

Figure \ref{fig:log-Mana} neatly highlights the intuition behind a potential attack strategy for this communication network. When $s$ is large and $\rho$ is small, the elongated shape taken by the network results in the creation of a choke point separating the few high Mana nodes from the many low Mana ones.

In an eclipse attack, the attacker's goal is to control all the victim's incoming and outgoing communications, isolating the target from the rest of the peers and destroying the ability of all peers to reach a consensus\cite{singh2006eclipse}. In our work, the goal of the attacker is to achieve control over all links separating two sub-components of the network: this allows the attacker to manipulate the flow of information between the two sub-components, gaining advantage and twisting consensus towards their needs. Since the parameters of the network formation protocol are known to the attacker, they only need to find an allocation of their Mana endowment across multiple Sybil nodes in a way that allows them to control the crucial choke points.

To measure the attack efficiency, we introduce the concept of damage $D$ and cost $x$. The damage is measured in terms of the fraction of Mana the attacker is able to disconnect from the network: if the total Mana in the system is denoted as $M = \sum_i m(i)$ and the network nodes are split into two components $A$ and $B$ where without loss of generality $\sum_{i \in A} m(i) \leq \sum_{j \in B} m(j)$, then 
\begin{equation}\label{q:3}
 D(A) = \sum_{i \in A} m(i)/M .
\end{equation}
Clearly, the maximum attainable damage is then $D_{max} = 1/2$.

To control a link, the attacker needs to control one of the endpoint nodes.
The cost of controlling a link $(i,j)$ is defined as the minimum Mana of its endpoint nodes $x(i,j) = \min (m(i), m(j))$. Then, for a set of links $C(A,B)$ which separates partition $A$ from $B$ and the set of least Mana endpoints (the ``frontier") $F = \lbrace \arg\min (m(i), m(j)) \; \forall (i,j) \in C(A,B) \rbrace$, the cost of attack reads 
\begin{eqnarray}\label{q:4}
x(C) = \sum_{i \in F} m(i) .
\end{eqnarray}

The goal of the attacker is, for a given Mana cost $x_{max}$ they are able to commit, to find the optimal set $C$ such that $x(C) \leq x_{max} $ and  $x(C') > x_{max} \forall C'$  such that  $D[x(C')] \geq D[x(C)]$.

This is a NP-hard problem, as there are exponentially many partitions and cuts from which to choose. In the following, we propose two simplified strategies for the attacker: a ``Betweenness" strategy and a ``Greedy" strategy, which both require full information about the P2P structure, and a derived ``Blind" strategy that only requires the more realistic assumption of knowledge of the parameters used to generate the network.

\subsection{Attack strategies}

\paragraph{Betweenness strategy}

The betweenness strategy, as the name suggests, relies on edge betweenness centrality to identify the optimal cut. Similar to node betweenness, edge betweenness serves as a metric that quantifies the fraction of shortest paths passing through a specific edge between any two nodes within a network \cite{freeman1977set}.
Inspired by the Girvan-Newman algorithm for community detection \cite{girvan2002community}, the strategy prescribes removing edges from the network starting from the ones with the highest betweenness centrality.
However, unlike the community detection process, in this scenario, the removal of the top betweenness links is halted once the network is divided into two parts. It is crucial to note that after each removal of the top betweenness link, the link betweenness is recalculated for the entire network. The attacker proceeds iteratively, removing the top betweenness links until the original network is fragmented into two distinct sub-networks, denoted as $A$ and $B$. The removed links constitute the associated cut $C$, and the attack damage $D$ is the total Mana associated to the nodes in the sub-network deriving from cut $C$, which has the smaller total Mana share.

\paragraph{Greedy strategy}
In the literature, we can find many different instances of network partitioning algorithms, such as spectral partitioning \cite{fiedler1973algebraic}, clustering coefficient \cite{saramaki2007generalizations}, community detection  \cite{yang2016comparative} or heuristic based partitioning~\cite{kernighan1970efficient}. However, none of these algorithms takes advantage of node attributes, such as the Mana endowment. Here, we introduce a simple method to split the auto-peering network according to the Mana ranking information and the network topological information, which we call the ``Greedy" strategy. 

The Greedy strategy is much more simplistic than the Betwenness one: the attacker splits the nodes into two sets $A = \lbrace i \leq i^* \rbrace$, and  $B = \lbrace i > i^* \rbrace$ for each possible target node $i^*=1,\dots N-1$. It follows that the cut is the set of links connecting $A$ and $B$.
For each $i^*$ we calculate the cost of the split and choose the $i^*$ that maximises damage per unit cost to run the attack.


\paragraph{Blind strategy}
In practical situations, the specific topology of the P2P network is not available, as nodes do not have a full picture of all connections in the network but only know their own peers. A blind strategy, one that does not require full knowledge about the network topology, is therefore needed. We design a blind strategy that is \textit{informed} by the results obtained with the Betweenness and Greedy strategies, namely measuring the frequency with which a specific node, given the network formation parameters, ends up being in the control set identified by the full-information strategy. We then take the most frequently targeted node $i_\sigma$, where $\sigma \in \lbrace B,G\rbrace$ (for Betweenness and Greedy, respectively) and define the control set for the strategy $\mathcal{C}_{\text{blind}}$ as all nodes within a range $L$ in the Mana ranking from node $i_\sigma$, i.e. 

\begin{equation}
    \mathcal{C}_{\text{blind}}(\sigma) = \left\lbrace i \; : \; \vert i_\sigma - i \vert < L \right\rbrace
\end{equation}

The attacker is assumed to take control of this set of nodes and we verify whether this was successful to split the network and quantify the attack efficiency according to our Damage and Cost measures.

\section{Results}

\subsection{Simulation results for full information attacks}

We explore the parameter space and find that the resulting network structure is most sensitive to the value of Zipf's exponent $s$, which describes Mana heterogeneity (see section \ref{sec:robustness}). We fix all other parameters to reasonable values $N=100$, $R=10$, $\rho=4$, $k=4$, and focus on the attack vulnerability varying $s$ from $0.5$ to $1.5$ at intervals of $0.1$. For each set of parameters, we generate a sample of 1000 networks and apply our attack strategies, measuring the damage over cost ratio $D/x$.

\begin{figure*}[t]
         \centering
         \includegraphics[ width=0.9\textwidth]{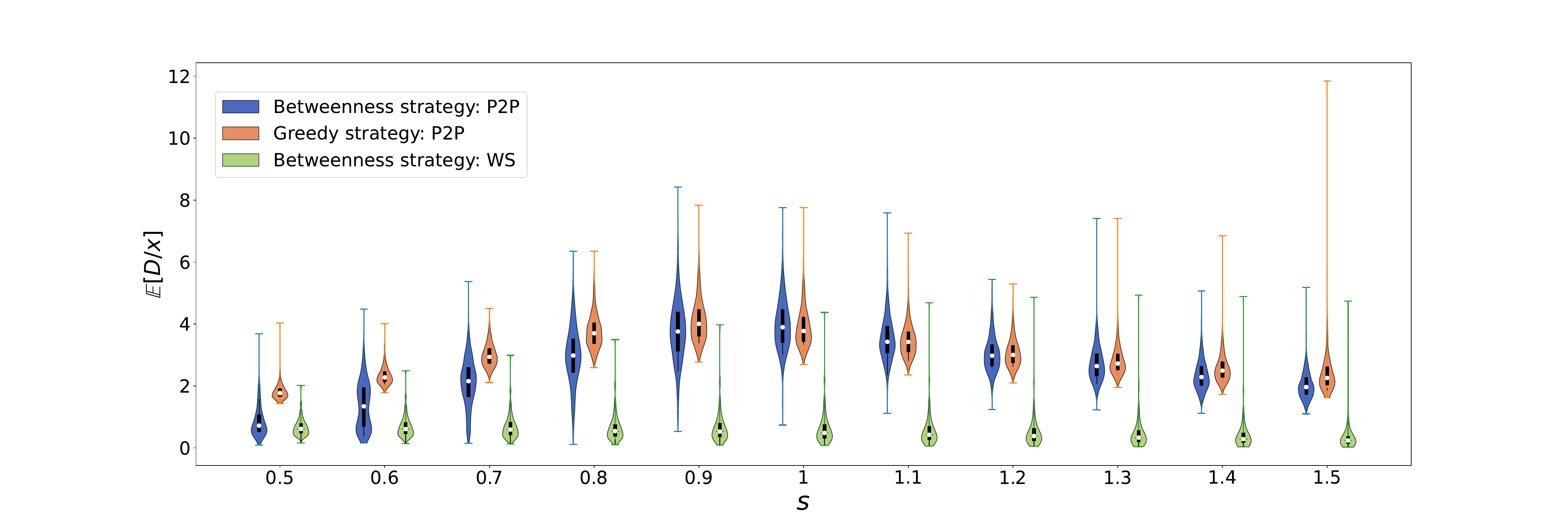}
    \caption{The average damage over cost ratio by attack strategy and underlying network formation model. Each auto-peering generated network has $N=100$, $\rho=4, R=10$, $k=4$. WS generated networks have $N=100, k=4, p=1$, the data size is 1000 graphs.}
    \label{fig: greedy-bt damage violin}
     \end{figure*}

In Figure \ref{fig: greedy-bt damage violin}, we present the results for the average damage over cost $\mathbb{E}[D/x]$ over 1000 simulations and across varying values of $s$ for the full information attacks. 
We observe that both the Betweenness and Greedy strategies, respectively represented by the blue and orange violins, attain maximum efficiency when $s \approx 1$. Additionally, we show the result of applying the Betweenness strategy to a random-regular network where Mana does not have a role in the formation of connections, i.e. the fully randomised Watts-Strogatz (WS) model of small-world networks \cite{watts1998collective} with coordination number $k$.
Compared to the auto-peering network formation model, the attack efficiency in the WS model appears to be much closer to 0 in the vast majority of cases. The result in Figure \ref{fig: greedy-bt damage violin} shows that in specific Mana distribution conditions, the attacker can achieve relatively large damage to the network with a small cost and that this vulnerability is induced by the presence of Mana in the network formation protocol. Both strategies achieve maximal $D/x \approx 3.5$ for $s=1$: this is particularly relevant because Zipf laws with exponent in the close range around $1$ have been observed to describe wealth distribution in cryptocurrency token holdings in the literature \cite{kusmierz2022centralized}.

Next, we compare the frequency of endpoint removals between the betweenness strategy and the greedy strategy under the conditions of $\rho=4$, $R=10$, and $N=100$, based on a sample of $1000$ graphs. Our findings, which for the sake of conciseness we report in  Fig.A2 and Fig.A3 in the Appendix, reveal that the greedy strategy, which aims to optimise the $D/x$ ratio, typically removes node $i_{G}=14$ as the maximum frequency endpoint. On the other hand, for the betweenness strategy, the maximum frequency endpoint that is removed is node $i_{B}=12$. 
We find that in the case of $s=1$, the endpoints selected with maximum frequency $F$ by both the ``betweenness" strategy and the ``greedy" strategy are typically in the same range, albeit not precisely the same nodes.
We are going to use the maximum frequency endpoints hereby identified to inform the blind attacks analysed in the next section.

\subsection{Simulation results for blind attacks}

\begin{figure*}[ht]
    \centering
    \includegraphics[width=0.8\textwidth]{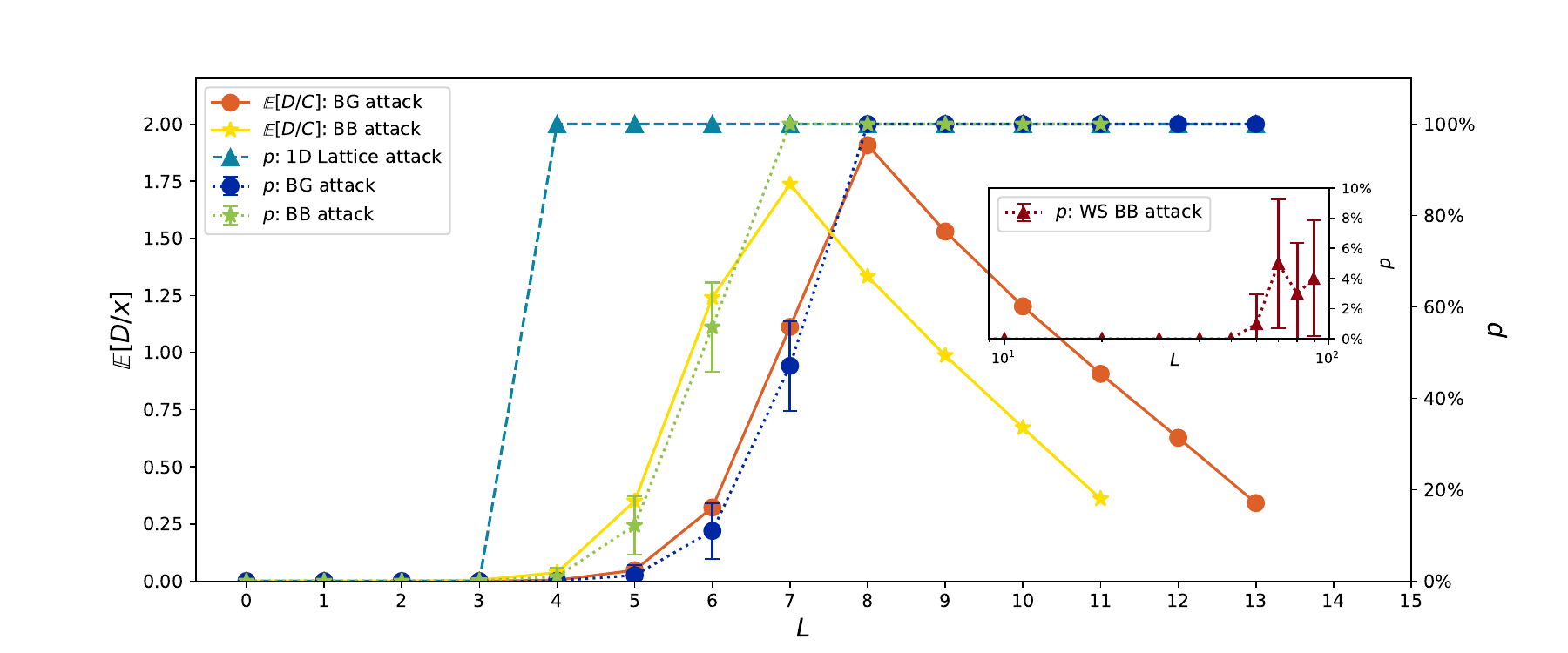}
    \caption{Success ratio $p$ (on the right $y$-axis) and average damage $\mathbb{E}[D/x]$ (on the left $y$-axis) per unit cost of a Blind strategy informed by the Greedy (``BG") and Betweenness (``BB") strategies. The results are averaged over a sample of $1000$ simulated networks with $N=100$, $\rho=4$, $R=10$ and $s=1$ while the error bars are $95\%$ confidence intervals. As comparisons, we also show the success ratio $p$ on an equivalent 1D lattice and a fully randomised regular WS-type network in the inset. 
    } 
    \label{fig:results-merged}
\end{figure*}

A pitfall of the two strategies above is that they both require the attacker to have perfect knowledge of the network structure, which is unrealistic. A more realistic relaxation of this assumption is that the attacker is only aware of the network formation model parameters. Given such information, the attacker can reproduce the above results and inform a ``Blind" strategy, aiming to control the nodes that are most often needed to perform the split. In particular, for the specific case of $N=100$, we take the ``target" nodes $i_B=12$ and $i_G=14$ (for the Betweenness and Greedy strategy respectively) and define the parameter $L$ such that the blind attacker controls nodes $\mathcal{C}_{\mathrm{blind}}(\sigma)$ for $\sigma \in \lbrace B,G \rbrace$. We then measure the $D/x$ ratio over $1000$ simulated networks along with the \textit{success ratio} $p$, i.e. the fraction of simulations where the network is successfully split by the blind attacker, and show the results in Figure \ref{fig:results-merged} as a function of $L$.


The red and yellow lines show $\mathbb{E}[D/x]$, the average damage over cost attack efficiency associated with a Blind strategy informed either by the Greedy (red, ``BG") or Betweenness (yellow, ``BB") results, averaged over a sample of $1000$ networks. The blue and green lines track the attack success rate $p$ using the target node from the ``Greedy" strategy (blue line) and from the ``Betweenness" strategy (green line).

As should be expected, for low choices of $L$ the strategy is unsuccessful in splitting the network. The Blind-Betweenness (BB) strategy reaches a peak average efficiency of $\mathbb{E}[D/x]=1.75$ for $L^*_{b}=7$, corresponding to $100\%$ attack success.
On the other hand, the Blind-Greedy (BG) strategy reaches peak efficiency for $L^*_{g}=8$: while requiring more nodes to be controlled, the efficiency is larger on average ($\mathbb{E}[D/x]=2$). Summarising, both strategies allow the attacker to successfully split the network from a low information context. Based on our simulations the BG strategy attains a better efficiency, but we observe that we did not assign an additional cost to the number of controlled nodes, we only considered the Mana required to control the node: the number of nodes to control may affect the cost of an attack in a real-world situation.
Interestingly, both strategies at their peak efficiency $L$ command the same cost of around $24\%$ of all the Mana in the system, as shown in Figure \ref{fig: control-percent}.
This means that an attacker controlling $24\%$ of the system's Mana would be able to identify a distribution strategy to control the choke points of the P2P network $100\%$ of the times.

To compare the severity of the attacks with some baseline scenarios, we also run the same attacks on a 1D lattice with the same coordination number $k$ as the auto-peering network and on a fully randomised Watts-Strogatz (WS) network model, i.e. a random-regular network of degree $2k$. For the one-dimensional lattice, for which we show the success ratio with the cyan triangles in Figure \ref{fig:results-merged}, it is straightforward to see that the network is always split whenever $L \geq k$, where in this case $k=4$. Compared with a 1D k-regular lattice, the auto-peering networks are more resilient.
This makes perfect sense: 1D lattice are completely predictable networks, where the attacker does not need to acquire information to predict the topology of the network.
On the other extreme the WS networks, being fully randomised, are much more resistant to attacks and the networks are almost never successfully split unless $L$ is chosen to be of the same order of magnitude as $N$. We show this result in the inset of Figure \ref{fig:results-merged}.


This result shows clearly that the IOTA auto-peering protocol produces networks that are in between the fully deterministic 1D lattice and a fully random regular network. Depending on the parameter values, the Mana-based connection rules can severely restrict the options for nodes to connect between themselves, resulting in the loss of the security provided by randomness and consequently in predictable network structures that can be exploited.

\subsection{Strategy Robustness}\label{sec:robustness}

In order to verify the robustness of our results, we test the proposed strategies by simulating the model on a realistic range in the parameters' space. We choose $\rho\in[1.5,4]$ and $s\in[0.5,1.5]$ as they should provide a comprehensive overview of realistic parameters choices. 


We find that, as long as $N$ is large enough compared to $k$, varying the number of nodes has no particular effect on the results. This is mostly due to the fact that the Mana is distributed according to Zipf's law, and so most of the Mana is concentrated in the first few nodes in the ranking regardless. Similarly, we do not find effects of particular interest when changing $R$ within reasonable ranges. For the sake of space, we report additional results testing $N$ and $R$ dependence in Fig.A5 and Fig.A6 (see Appendix). For each combination of parameters, the results were averaged over a sample of $1000$ graphs generated by the auto-peering network formation model.
On the other hand, we find an interesting interplay between the $\rho$ and $s$ parameters, which we discuss in the present section.



\begin{figure}[t]
    \centering
 \includegraphics[width=0.45\textwidth]{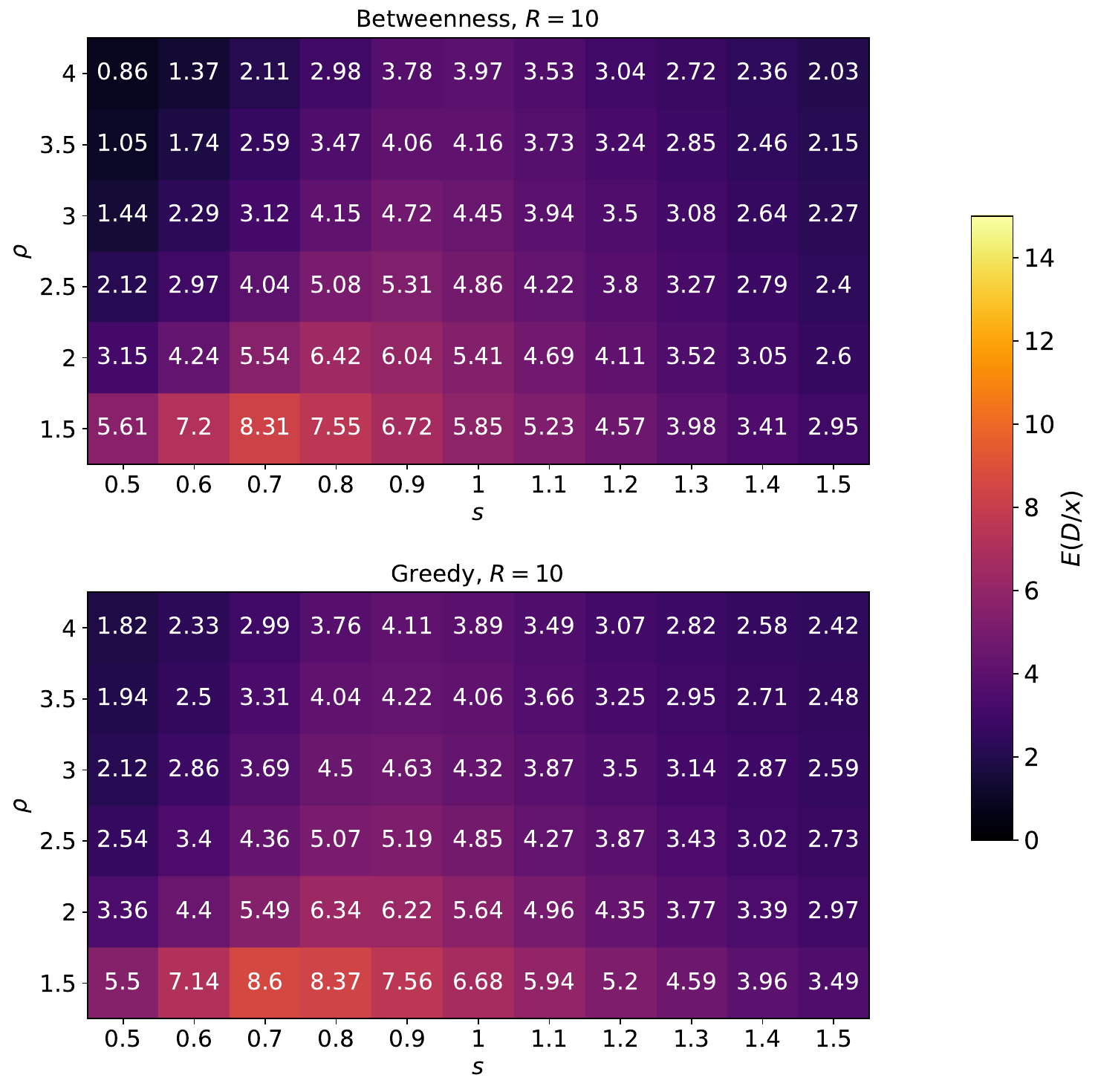}
    \caption{Heatmaps of the expected damage over cost, $\mathbb{E}[D/x]$, plotted over  $\rho$ and $s$, for the betweenness (up plot) and greedy (down plot) strategy. $R=10$ is kept constant. The maximum for each fixed $s$ is when $\rho$ is minimal, meaning that whenever the P2P model is more similar to a chain, the network is maximally vulnerable. } 
    \label{fig: bt-gr-DC-overview}
\end{figure}

In Figure \ref{fig: bt-gr-DC-overview} we show the expected damage over cost ratio across different parameter combinations for the two informed strategies. While the performances of the two strategies are slightly different, we do observe a similar trend: both attain maximum success (in terms of damage over cost) for low $\rho$ and for $s$ between $0.5$ and $1$, i.e. when the Mana distribution is sufficiently heterogeneous compared to nodes' ``tolerance" $\rho$, but not so heterogeneous that attacks can only be successful by controlling the top Mana nodes themselves. At the same time, we observe that $\rho$ has a similar effect on both strategies for all values of $s$, with a stronger effect when $s$ is low (so when Mana distribution is homogeneous). This is explained by the role of $\rho$, which is increasing (by a multiplicative factor) the tolerance for each node in terms of Mana difference to its potential neighbours; this means that when the network is homogeneous, a large $\rho$ implies longer range connections, while when $s$ is very large, the randomness effect of $\rho$ is diminished. 


\begin{figure}[t]
    \centering \includegraphics[width=0.5\textwidth]{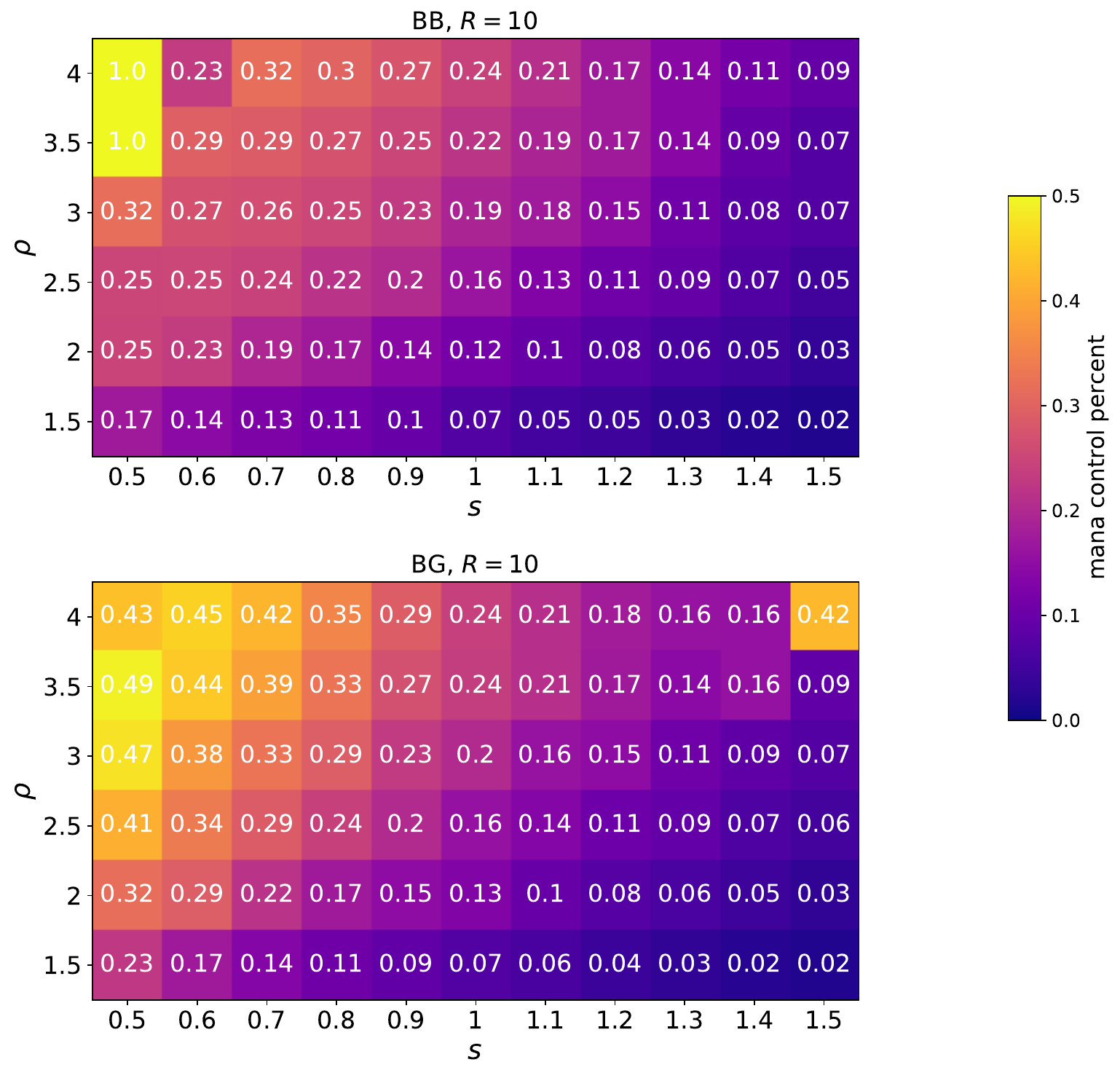}

    \caption{Cost in Mana (as a percentage of the total Mana in the system) of a blind attack strategy inspired by betweenness strategy results (bottom plot) and greedy strategy results (upper plot) necessary to obtain a successful network split $100\%$ of the times, the results are averaged over $1000$ simulations for each parameters combination. $s$ and $\rho$ vary, while $R=10$ and $N=100$ are kept constant.} 
    \label{fig: control-percent}
\end{figure}

In Figure \ref{fig: control-percent} we present the Mana cost (as a percentage of the total Mana) that is necessary to successfully split the network with $100\%$ success, following a blind attack with the BB or BG strategy. We notice that the regions of the parameters space which deliver the worst output in terms of attack efficiency are also the ones with the largest cost of attack, where graphs generated form the auto-peering model resemble random regular graphs. The cost is monotonically decreasing in $s$ and increasing in $\rho$ for both the blind strategies. 


\section{Conclusions}

In conclusion, our results point out that a malicious agent controlling just over $24\%$ of the total Mana can divide their Mana into a set of nodes that will most likely act as choke points of the P2P network from the auto-peering network formation model, leading to the attacker successfully splitting the P2P network and achieving damage that is close to the maximum possible. Such an attack would potentially end up compromising the operations of the IOTA network, as the attacker can control the flow of information between the high-Mana portion of the network and the vast majority of nodes on the other side. 
For comparison, we also apply the strategies on 1D lattices and fully rewired Watts-Strogatz (WS) networks, finding that the IOTA auto-peering protocol is far less resistant than the latter while slightly more resistant than the 1D lattice.

To conclude, we want to provide a practical comment to interpret our results: the auto-peering network formation model is not currently implemented on the IOTA system; as such there is no immediate risk to the IOTA network.

The results of the present work are to be interpreted as a policy suggestion, presenting a potential attack vector. At the same time, from a network science perspective, we believe that the auto-peering model provides an interesting interpretation of random assortative formation models, and provides an alternative bridge between k-regular Poisson graphs and 1D lattices, complementing the popular Watts-Strogatz model.

\section*{Acknowledgments}
YG acknowledges financial support from the China Scholarship Council. AST acknowledges support by FCT through the LASIGE Research Unit, ref. UIDB/00408/2020 (https://doi.org/10.54499/UIDB/00408/2020) and ref. UIDP/00408/2020 (https://doi.org/10.54499/UIDP/00408/2020).



\bibliographystyle{IEEEtran}
\bibliography{bibliography}
\newpage
~
\newpage

{\appendix
In this appendix, for a better understanding of the strategies employed in the paper, we present additional details regarding the simulation results of the two full information strategies (``Betweenness strategy" and ``Greedy strategy"), which in turn were used to develop the ``Blind strategy."

In Figure \ref{fig: greedy damage over cost}, we show the simulation results of $E[D/x]$ for $1000$ graphs, where $N=100$, $R=10$, $\rho=1$, and $N=100$, when applying the Greedy strategy. This figure explains the target selection mechanism of the Greedy strategy, which selects the target node $i^*$ based on the highest $E[D/x]$ value among the $N-1$ potential splits.

We then compare the consistency of the Betweenness strategy and Greedy strategy in terms of the nodes that are typically included in the control set. Under the same parameters conditions used to produce Figure \ref{fig: greedy damage over cost}, our findings in Figure \ref{fig:btVSgreedy} reveal that the greedy strategy most often targets node $i_G=14$, whereas the betweenness strategy's most frequently targeted node is $i_B = 12$. 

To compare the attack efficiency in these two strategies, we also reframe the result shown in Figure 4 of the main text by showing the ratio between the two $E[D/x]$. The result is displayed in Figure \ref{fig: bt-over-greedy-damage-over-cost}, which shows that while for the large majority of parameters the two strategies are similarly effective, in the top-left corner (when $\rho$ is large and $s$ is small) the greedy strategy significantly outperforms the betweenness strategy. This is the same region of the parameters space where both strategies have the worst efficiency, as reported in Figure 4 of the main text. It is also the range of parameters where the networks produced by the auto-peering model are maximally close to a random-regular topology.
In Erdos-Renyi graphs the betweenness information is homogeneous and the attacker gains no advantage from the distribution of link betweenness.

Figure \ref{fig: blind-heat} provides an overview of the effect of BB (Blind-Betweenness) and BG (Blind-Greedy) attacks. Here the heatmap indicates the minimum $L$, that is the count of nodes removed before and after the target nodes $i_\sigma$ identified by the full-information strategies, such that the blind attack is $100\%$ successful in splitting the simulated graphs. This measure can be intended as a measure for the robustness of the generated networks. We see that in BB attacks the value of $L$ is relatively stable until it explodes in the region of the parameters' space where the network is more akin to an unstructured random regular network, i.e. low $s$ and high $\rho$. The $L$ values of BG attacks are more gradually increasing when moving towards the same region of the parameters' space. The overall take-home message about the robustness of the different methodologies is that BG attacks are less sensitive to parameter variation.

Finally, we provide a sensitivity analysis for our results with respect to the parameters $R$ and $N$ in Figures \ref{fig: bt-over-greedy-vary-N} and \ref{fig: greedy damage over cost vary R}. The analysis demonstrates that both network size $N$ and parameter $R$ present minimal quantitative influence on the results, and do not affect the overall qualitative behaviour. In Figure \ref{fig: bt-over-greedy-vary-N}, varying $N$ yields consistently marginal differences in $E[D/x]$ across the subplots. Similarly, Figure \ref{fig: greedy damage over cost vary R} indicates that altering $R$ produces negligible variations in $E[D/x]$ within each subplot grid. These observations confirm the robustness of the outcomes to changes in both network size $N$ and the parameter $R$ in the additional peer discovery condition.
\vspace{-0.5cm}
\begin{figure}[htbp]
         \includegraphics[ width=0.5\textwidth]{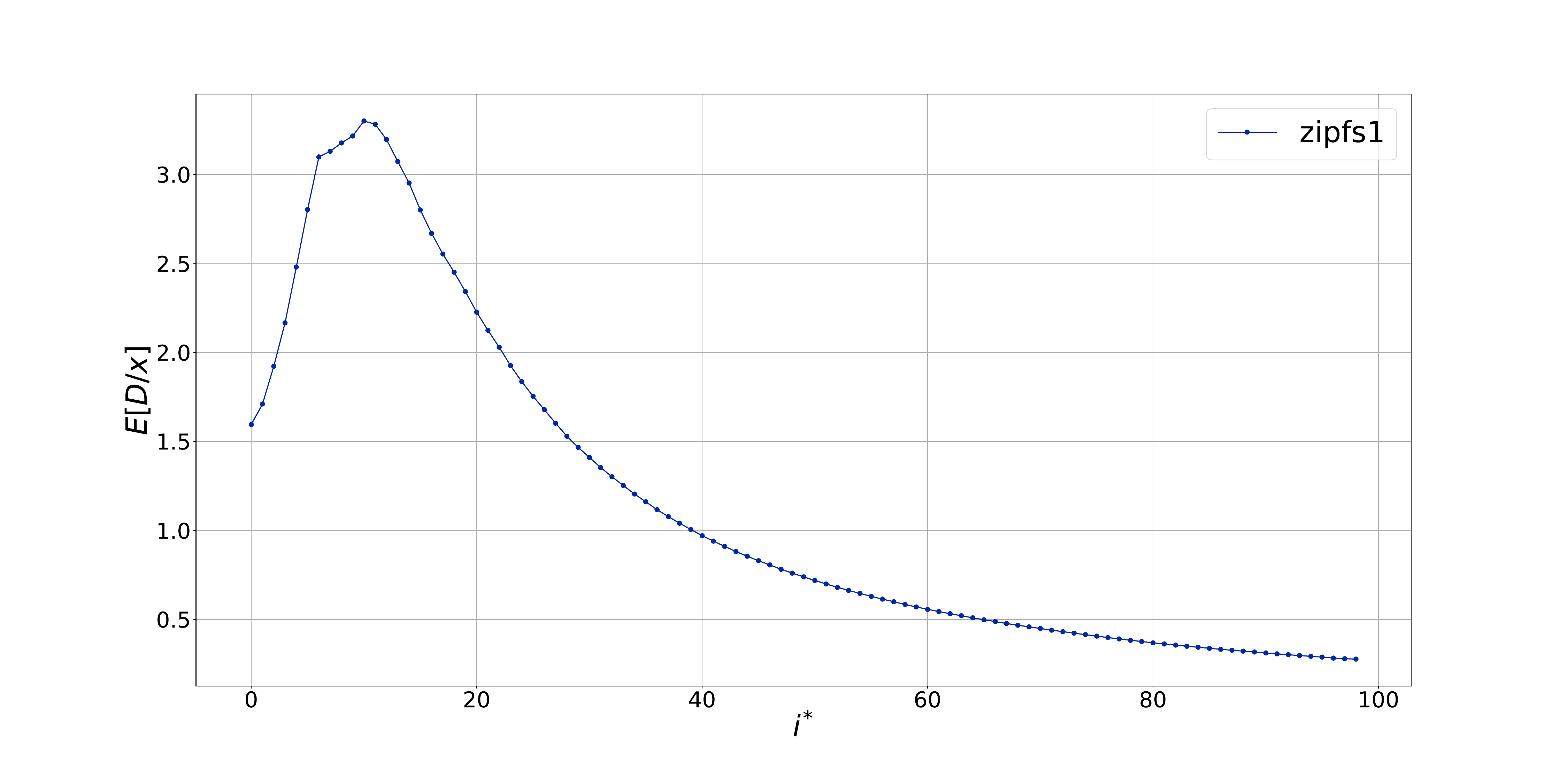}
         \renewcommand{\thefigure}{A1}
    \caption{The average Greedy damage over cost by targeted node $i^*=1,\dots N-1$, where N=100, $\rho=4, R=10$, and sample size is $1000$ graphs.}
    \label{fig: greedy damage over cost}
     \end{figure}


\begin{figure}[htbp]
    \centering
    \includegraphics[width=0.5\textwidth]{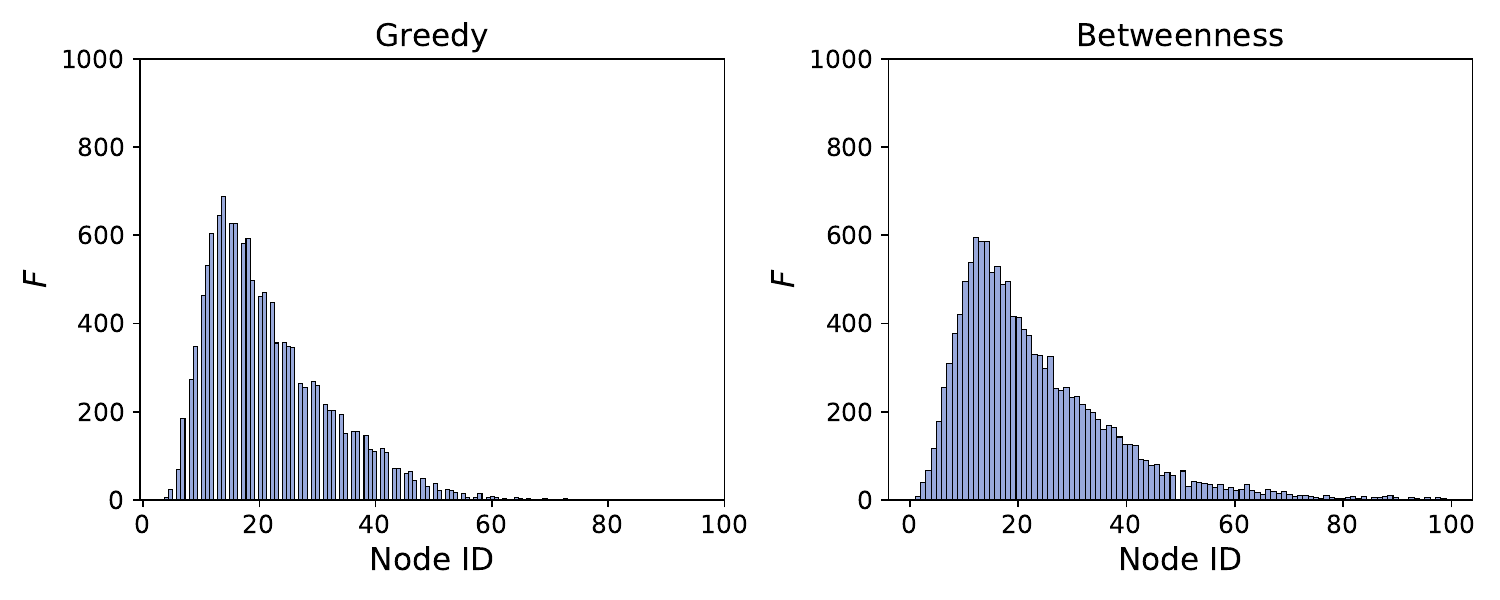}
    \renewcommand{\thefigure}{A2}
    \caption{Histogram of frequencies $F(i)$ with which node $i$ is in the control set according to the Greedy and Betweenness strategies on a sample of $1000$ graphs, where $\rho=4, R=10, N=100$, $k=4$, $s=1$. } 
    \label{fig:btVSgreedy}
\end{figure}
\begin{figure}[htbp]
    \centering \includegraphics[width=0.5\textwidth]{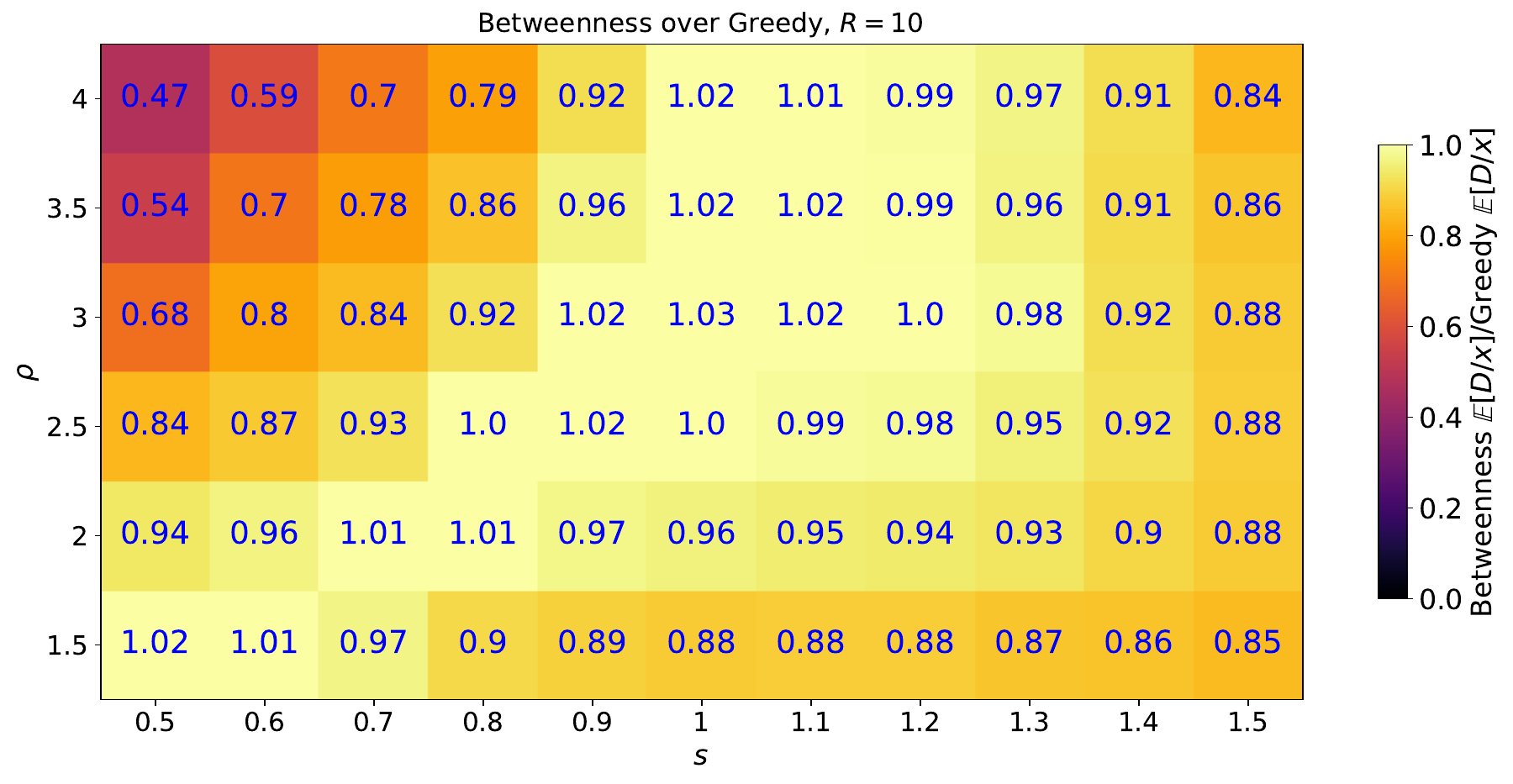}
    \renewcommand{\thefigure}{A3}
    \caption{The heatmap shows the ratio of $\mathbb{E}[D/x]$ for the Betweenness strategy over the $\mathbb{E}(D/x)$ for the Greedy strategy, effectively representing the relative efficiency of the betwenness attack with respect to the greedy strategy. $s$ and $\rho$ varies, while $R=10$ is kept constant. The reader can notice a minimum placed at $s=0.5$ and $\rho=3.5$. That is the region of the parameters' space where the auto-peering model networks most resemble regular Erdos-Renyi random graphs, and the link betweenness becomes homogeneous and highly uninformative.} 
    \label{fig: bt-over-greedy-damage-over-cost}
\end{figure}

\begin{figure}[htp]
    \centering \includegraphics[width=0.5\textwidth]{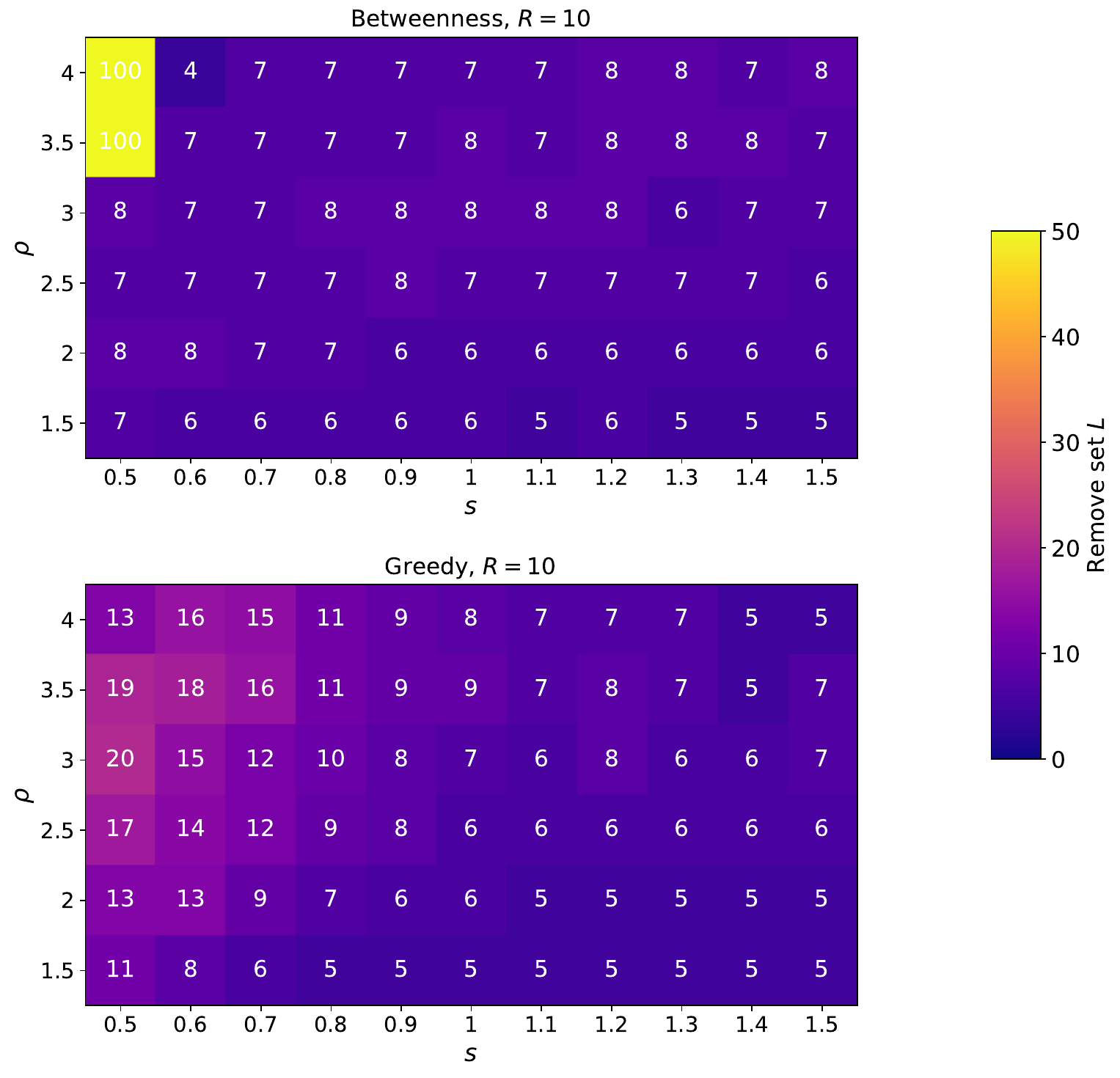}
    \renewcommand{\thefigure}{A4}
    \caption{Each cell's number indicates the minimum value of $L$ necessary to obtain a successful network split $100\%$ of the times in the betweenness and greedy strategies respectively. The results are averaged over $1000$ simulations for each parameters combination. $s$ and $\rho$ vary, while $R=10$ is kept constant.}
    \label{fig: blind-heat}
\end{figure}

\begin{figure}[htp]
    \centering \includegraphics[width=0.5\textwidth]{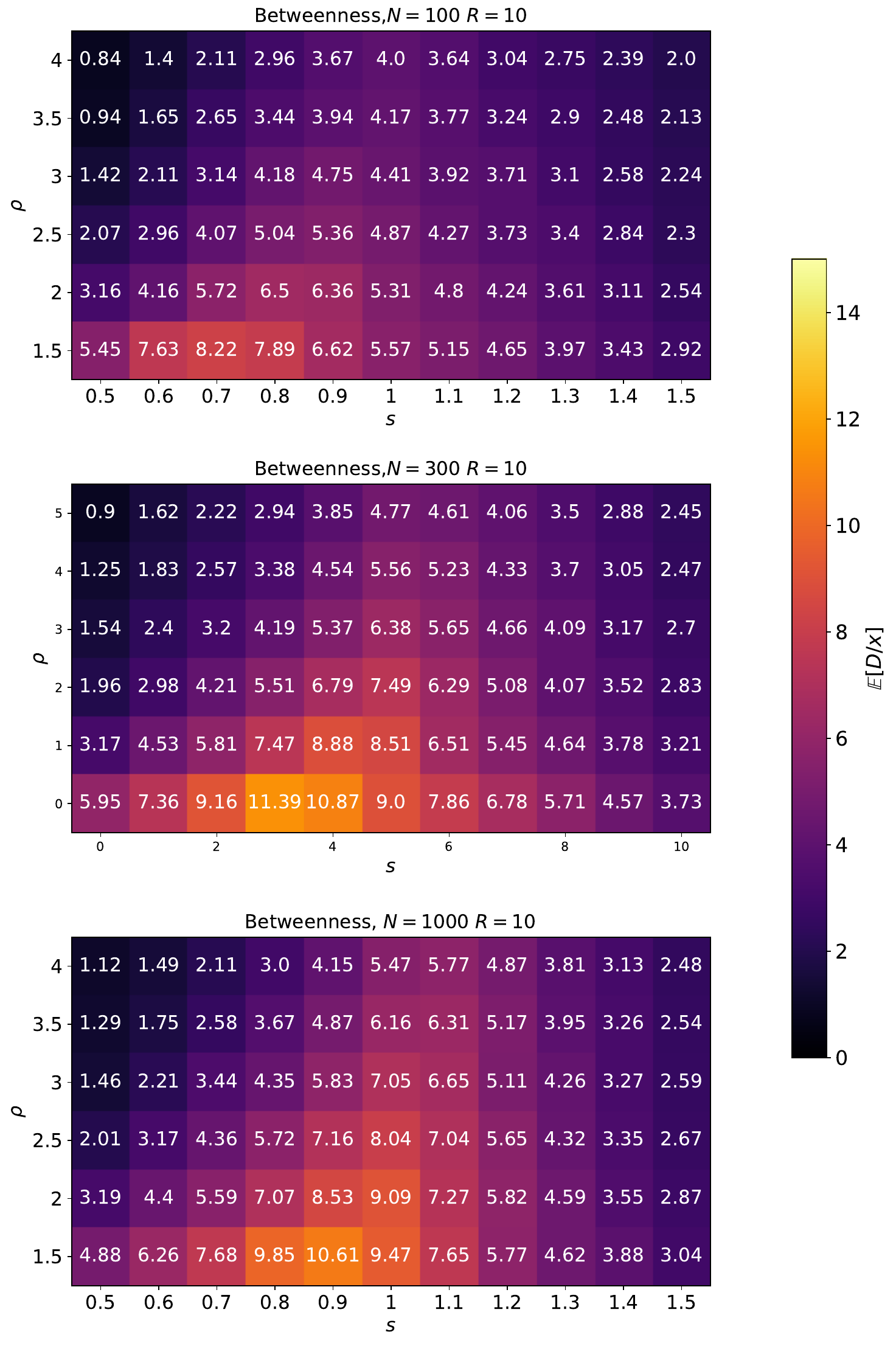}
    \renewcommand{\thefigure}{A5}
    \caption{The heatmap shows the ratio of $\mathbb{E}[D/x]$ for the Betweenness strategy $\mathbb{E}(D/x)$, where $s$, $N$ and $\rho$ varies, while $R=10$ is kept constant. From the three heat map plots, we can see that there is no big gap for the damage over cost in different network sizes.} 
    \label{fig: bt-over-greedy-vary-N}
\end{figure}


\begin{figure}[htp]
         \includegraphics[ width=0.5\textwidth]{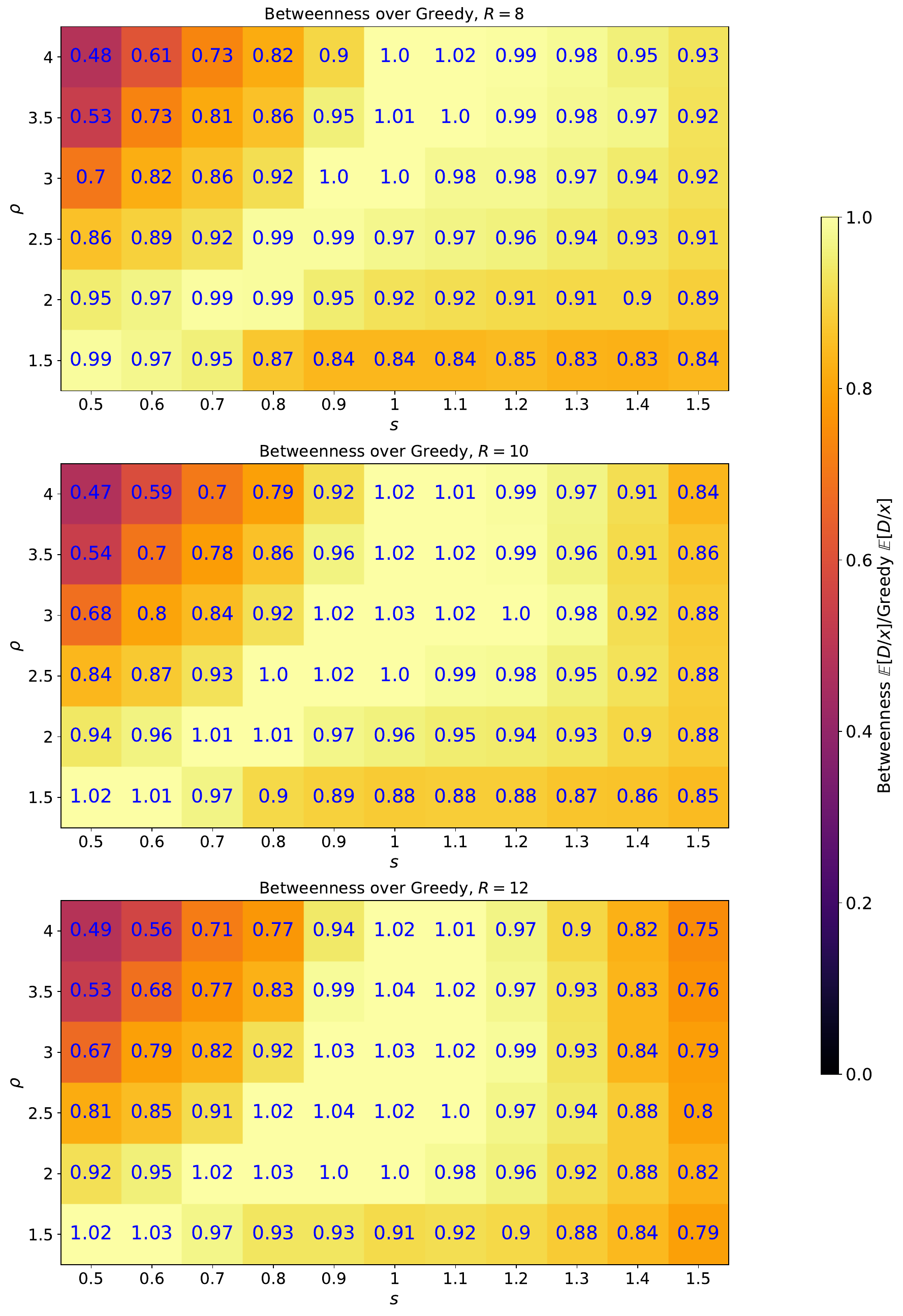}
         \renewcommand{\thefigure}{A6}
    \caption{The heatmap shows the ratio of $\mathbb{E}[D/x]$ for the Betweenness strategy over the $\mathbb{E}(D/x)$ for the Greedy strategy varying $s$, $R$ and $\rho$, where $N=100$. }
    \label{fig: greedy damage over cost vary R}
     \end{figure}}

\end{document}